\documentclass[conference]{IEEEtran}
\IEEEoverridecommandlockouts

\usepackage{cite}
\usepackage{amsmath,amssymb,amsfonts}
\usepackage{algorithmic}
\usepackage{graphicx}
\usepackage{textcomp}
\usepackage{xcolor}
\def\BibTeX{{\rm B\kern-.05em{\sc i\kern-.025em b}\kern-.08em
    T\kern-.1667em\lower.7ex\hbox{E}\kern-.125emX}}

\usepackage{pgfplots}
\usepackage{subcaption}
\usepackage{graphicx}
\usepackage{amsmath}

\begin{document}
\title{Situational Awareness for Safe and Robust Multi-Agent Interactions Under Uncertainty}


\author{\IEEEauthorblockN{Benjamin Alcorn\IEEEauthorrefmark{1}, Eman Hammad\IEEEauthorrefmark{1}\IEEEauthorrefmark{2}}
\IEEEauthorblockA{\IEEEauthorrefmark{1}Department of Electrical \& Computer Engineering\\}
\IEEEauthorblockA{\IEEEauthorrefmark{2}Department of Engineering Technology \& Industrial Distribution\\
Texas A\&M University, College Station, TX}
}

\IEEEoverridecommandlockouts
\IEEEpubid{\makebox[\columnwidth]{978-1-5386-5541-2/18/\$31.00~\copyright2025 IEEE \hfill} \hspace{\columnsep}\makebox[\columnwidth]{ }}

\maketitle

\IEEEpubidadjcol

\begin{abstract}
    Multi-agent systems are prevalent in a wide range of domains including power systems, vehicular networks, and robotics. Two important problems to solve in these types of systems are how the intentions of non-coordinating agents can be determined to predict future behavior and how the agents can achieve their objectives under resource constraints without significantly sacrificing performance. To study this, we develop a model where an autonomous agent observes the environment within a safety radius of observation, determines the state of a surrounding agent of interest (within the observation radius), estimates future actions to be taken, and acts in an optimal way. In the absence of observations, agents are able to utilize an estimation algorithm to predict the future actions of other agents based on historical trajectory. The use of the proposed estimation algorithm introduces uncertainty, which is managed via risk analysis. The proposed approach in this study is validated using two different learning-based decision making frameworks: reinforcement learning and game theoretic algorithms. 
\end{abstract}

\section{Introduction}
\label{sec:intro}
    Multi-agent systems (MAS) model multiple intelligent agents working to achieve common goals, with applications in fields such as autonomous vehicles, manufacturing, and robotics. They often operate with limited resources in uncertain environments, making situational awareness crucial. Agents with high autonomy but poor awareness may face performance, safety, or security issues \cite{zhang2023perception, 9827169}\cite{shimizu2021evaluation}. Thus, MAS helps provide a framework to help investigate and prioritize the development of safe and reliable systems \cite{9616449, BURTON2020103201}. Accurate environmental assessments improve navigation under uncertainty, and estimation algorithms help predict future states over time~\cite{de2020pysindy}. Estimation also ensures predictions remain reliable if the risk stays within a pre-defined limit \cite{bhadriraju2021risk}. 

Approaches like reinforcement learning struggle with accuracy and efficiency in complex setups \cite{rosolia2018data, troullinos2021collaborative}. Hence, it is essential to develop a framework balancing decision-making, awareness, and performance, despite resource limits \cite{chen2020deep} and decision delays \cite{homberger2007multi}.

For estimation, model-based methods create dynamic models of the environment. Model predictive control (MPC) optimizes future state costs \cite{EATON1992705, 9180048}, but it's resource-intensive \cite{holkar2010overview}, pushing focus towards data-driven approaches that train neural networks on trends. A more cost-efficient alternative is data-driven MPC \cite{9361343}, which does not require the demand of building and using a model of the system. Hence, a balance between model-based and data-driven approaches could be promising to achieve performance with constrained resources.

This work addresses a resource-efficient situational awareness approach handling uncertainties via constraining the autonomous agent to a set radius of observation and using neural networks (NN) based estimation in the absence of the environment observations. The main premise of the proposed approach is to limit resource consumption by limiting observability and action space used in the decision making process while adhering to safety guidelines. A basic model with two uncoordinated agents on a 2D grid is adopted to illustrate the benefits of the proposed approach with simplified dynamics. Two learning algorithms are evaluated under different conditions using performance and resource usage metrics. The proposed approach is useful for autonomous agents to be able to achieve results within set safety guidelines for themselves acting in the environment. The approach was observed to reduce safety violations for the overall environment as well. 

The key contributions of this work are the following: (1) develop a resource-efficient situational-awareness framework for autonomous agents via safety-based radius of observation, (2) integrate a learning-based estimation algorithm focusing on accuracy and efficiency to manage uncertainty; (3) provide a risk analysis of the estimation algorithm to support risk-informed operation, and (4) illustrate benefits of the proposed safety-oriented situational-awareness framework using two learning algorithms in a 2D basic dynamics case study.

\vspace{-0.1in}

\section{System Model}
\label{sec:sys-model}
This study demonstrates how situationally-aware agents achieve objectives under resource constraints and limited observability using a basic-dynamics model. Although it doesn't fully replicate real autonomous vehicle dynamics, it captures essential aspects for algorithmic purposes. Agents observe their environment and make decisions despite uncertainties like obstructions and noise. This work highlights that these principles can extend to complex environments including vehicle dynamics specifics like mass and acceleration. 

\begin{figure*}
    \centering
    \includegraphics[width=0.85\linewidth]{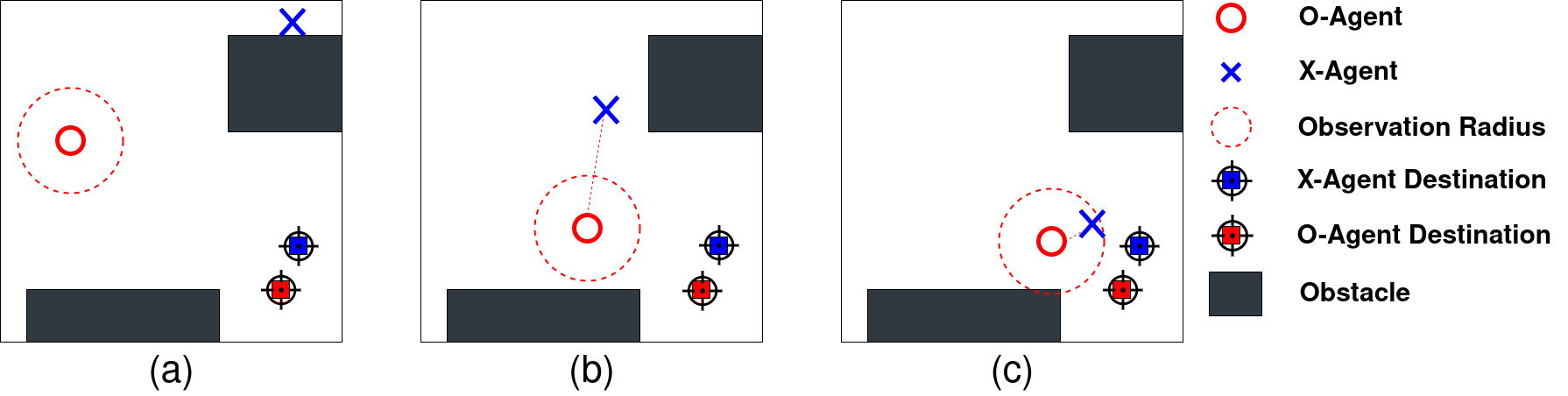}
    \caption{Example simulation states: (a) state where x-agent is not observable by the o-agent, (b) state where o-agent can detect that there is no x-agent within the observation radius, (c) state where o-agent detects x-agent within the observation radius.}
    \label{fig:environment}
\end{figure*}

The setup involves a grid of size $L \times L$ units where agents move from a random start to a random destination one step left, right, up, or down at each time step. Obstacles in the grid may cause collisions or block line of sight observations. We introduce the concept of the \textbf{observation radius} which is a uniform area around the agent where events such as obstacle or movements by another non-coordinating agent (e.g. x-agent) require attention and action. 

Figure \ref{fig:environment} depicts three scenarios that are encountered in the simulation: the initial state where the o-agent does not have any observations of other agents in the environment. Next, a state where the o-agent observes that the x-agent is not within the observation radius and thus acts based on initial optimal strategy, and a state where the o-agent detects the x-agent within the observation radius and have to follow the proposed algorithm to navigate. Ihe depicted case in this figure captures a context where is the observation radius is set as a fixed parameter. However, future work will extend this to build a strategy for an adaptive observation radius that varies based on factors reflecting safety risk tolerance such as uncertainty, noise, and perceived risk.


Assumptions are established in the environment to ensure practical constraints, such ensuring agents do not stagnate and would remain active within grid boundaries maintaining enclosed monitoring akin to road driving restrictions. Randomly assigned routes are evaluated for goal reachability and if if found impossible they are discarded which is crucial for learning algorithms. The study explores 1 - 1 agent scenarios to test validity. Future work will investigate scalability involving 1 - n and m-n interactions, plus controlled simulations to assess different target strategies and path limitations.

\section{Problem Formulation}
\label{sec-approach}
    \subsection{Approach}
\begin{figure}
    \centering
    \includegraphics[width=\linewidth]{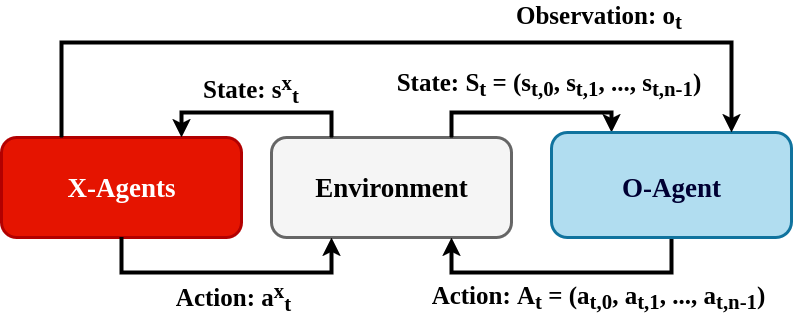}
    \caption{Diagram of the workflow of an environment of interacting agents with o-agents and x-agents that can only be interacted with through observations}
    \label{fig:flow}
\end{figure}
To study the ideas outlined, we designed a simulated autonomous vehicle environment in which agents interact under varying settings. One agent, referred to as the \textbf{o-agent}, acts as a decision-making vehicle that learns over time to adapt its behavior in the presence of other vehicles, known as \textbf{x-agents}, which follow a fixed, non-adaptive policy.

The o-agent uses a combination of real-time estimation, strategic adaptation, and dynamic observation to reach its destination efficiently and safely. Key design components include:

\subsubsection*{Observation Radius}

The o-agent selects a limited distance from itself to observe (its \textit{observation radius}) at any given time. This constraint mimics real-world limitations on sensing and processing along with giving control over the amount of area that the sensing covers to help restrict excess resource consumption. The agent can set the radius to be adaptive and influenced by measured confidence and risk in the environment, but this work focuses on a fixed radius that is set as a parameter for the simulation environment. A fixed radius will keep the radius constant throughout the entire trajectory planning sequence.

\subsubsection*{Estimation}
The o-agent leverages an estimator to predict the future actions of selected x-agents. A strategy is to predict the next \( n \) actions and update the estimate only every \( n \) time steps, trading off accuracy for reduced computational overhead. The estimate is constructed from a recurrent neural network (RNN) trained on historical state-action sequences of x-agents. This model maps a time series of observations to a distribution over future actions:
\begin{equation}
    f_\theta: (s_{t-k}^{x}, \dots, s_{t}^{x}) \mapsto \mathbb{P}(a_{t+1}^x)
\end{equation}

where \( \theta \) represents learned parameters.

The o-agent uses both point predictions and uncertainty estimates to decide when to rely on the estimator. To help establish the reliability of the estimator, Section~\ref{subsec:estimation} focuses on how many future time steps the agent can rely on the estimator for. 

\subsubsection*{Learning and Control}
At each step, the o-agent observes within its observation radius and determines if the x-agent is present. If it is not within the radius, then an optimal navigation algorithm, here Dijkstra's algorithm, will be used to find the strategy with the minimum Manhattan distance to the destination. In addition, if there is recent observation data from when the x-agent was within the observation radius and visible, then estimation will be performed to anticipate potential future interactions. This can be used when visibility is obstructed by an obstacle, as the estimator can give an accurate prediction for where the x-agent is, so the o-agent can act accordingly. 

Conversely, if the x-agent is observed within the observation radius, then a learning algorithm will be used. The agent can be configured to use either a game theoretic strategy or RL, both of which are outlined in Section \ref{sec:strats}.

\begin{figure}[h]
\centering
\includegraphics[width=0.47\textwidth]{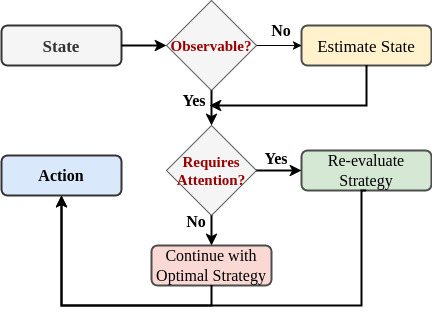}
\caption{Logical flowchart for decision-making in a situationally-aware agent}
\label{fig:logic-flow}
\end{figure}

This logical flow of the o-agent is outlined in Figure \ref{fig:logic-flow}.

\subsection{Objectives and Single-Agent Framework}

The o-agent operates in a discrete 2D environment and aims to reach a predefined destination while navigating safely among other agents. Hence, the agents key objectives are to (1) reach the goal in the fewest number of steps, (2) avoid collisions with x-agents and static obstacles, and (3) minimize unnecessary detours or route changes, all with minimal resources.

\subsubsection*{Control Setting}  
The o-agent has full control over its own state, represented as \( s_t^o \), and selects actions \( a_t^o \in A \) at each time step. The environment transitions according to:
\begin{equation}
s_{t+1}^o = f(s_t^o, a_t^o)
\end{equation}

\subsubsection*{Opponent Modeling}  
The x-agents are modeled as semi-deterministic agents that follow fixed policies. Their states are denoted as \( s_t^{x_i} \), and their actions are not influenced by the o-agent for this study.

\subsection{State and Action Spaces}
The state of the o-agent is composed of its observations within the observation radius $r_{obs}^o$ and the $(x,y)$ distance to the destination from its current position, represented as $d_{rel}$. To represent the observations within the observation radius, we use a matrix $M_t^o$ to show the state of surrounding positions of the o-agent when it is in position $(\alpha_t^o, \beta_t^o)$ at time $t$. This matrix is defined as:
\begin{equation}
    M_t^o = \begin{bmatrix}
        m^{\alpha_i-r, \beta_i-r} &  \dots & m^{\alpha_i-r, \beta_i+r}\\
        m^{\alpha_i-(r-1), \beta_i-r} & \dots & m^{\alpha_i-(r-1), \beta_i+r}\\
        \vdots & \ddots & \vdots\\
        m^{\alpha_i+r, \beta_i+r} & \dots & m^{\alpha_i+r, \beta_i+r}
        \end{bmatrix}
\end{equation}

Each of the values in the matrix take the following values, where spaces are either regarded as empty or occupied because all collisions, regardless of whether with another agent or with an obstacle, are treated the same. 

\begin{equation}
    m^{\alpha,\beta} = \begin{cases}
        0 & \text{$(\alpha, \beta)$ is empty} \\
        1 & \text{$(\alpha, \beta)$ is occupied}
    \end{cases}
\end{equation}

We define the state of the o-agent at time $t$ as $s^o_t \in \mathbb{R}^{(r_{obs}^o)^2+1}$ and is defined as:
\begin{multline}
    s_t^o = (m^{\alpha_o-r_{obs}^o, \beta_o-r_{obs}}, ..., m^{\alpha_o-r_{obs}^o,\beta_o+r_{obs}^o}, \\ m^{\alpha_o-r_{obs}^o+1,\beta_o-r_{obs}^o}, ..., m^{\alpha_o+r_{obs}^o,\beta_o+r_{obs}^o}, d_{rel})
\end{multline}

In each state, there are four possible actions that can be taken. This maps to a mathematical model where integers are used to represent each direction, such as 

$A_t^o = \{0,1,2,3\} = \{\text{up}, \text{left}, \text{down}, \text{right}\}$.

\subsection{Learning Strategies}
\begin{figure*}
    \centering
    \begin{subfigure}{0.42\textwidth}
        \centering
        \includegraphics[width=\linewidth]{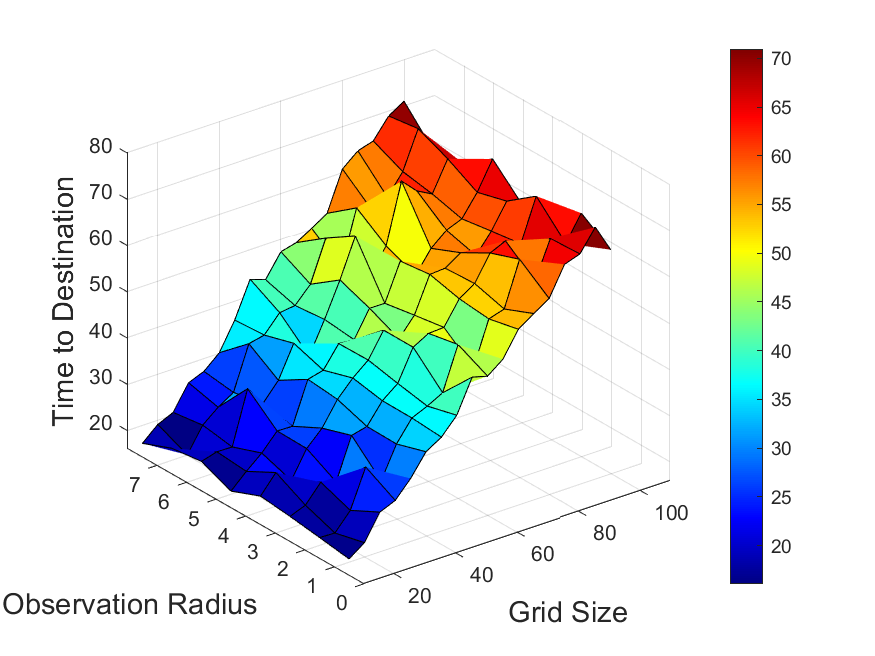}
        \caption{}
        \label{fig:image1}
    \end{subfigure}
    \hfill
    \begin{subfigure}{0.42\textwidth}
        \centering
        \includegraphics[width=\linewidth]{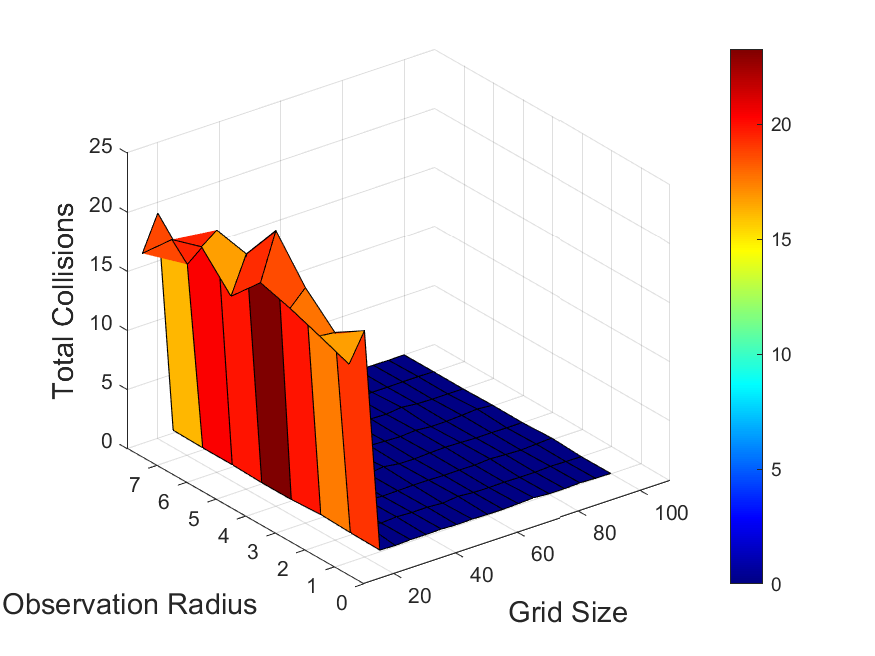}
        \caption{}
        \label{fig:image2}
    \end{subfigure}

    \vskip\baselineskip
    \begin{subfigure}{0.42\textwidth}
        \centering
        \includegraphics[width=\linewidth]{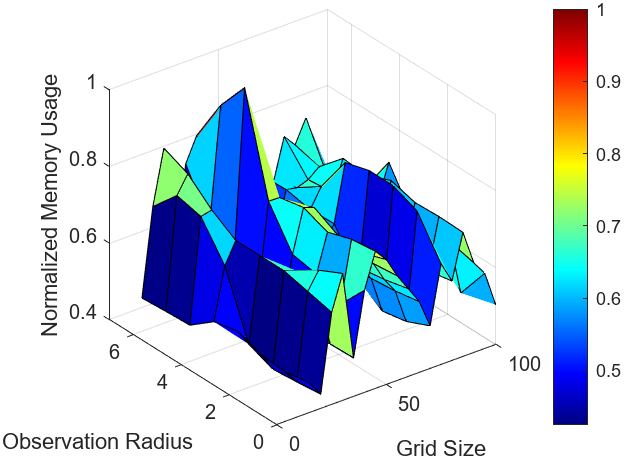}
        \caption{}
        \label{fig:image3}
    \end{subfigure}
    \hfill
    \begin{subfigure}{0.42\textwidth}
        \centering
        \includegraphics[width=\linewidth]{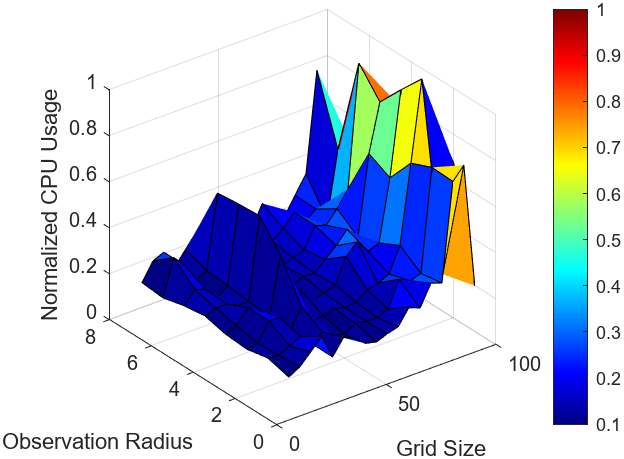}
        \caption{}
        \label{fig:image4}
    \end{subfigure}

    \caption{Comparison of performance metrics and resource utilization as a function of grid size ($L$) and observation radius $(r_{obs})$ using reinforcement learning. (a) Time to destination (unit time steps), (b) Total collisions out of 500 runs, (c) Memory usage (normalized percentage), (d) CPU Usage (normalized percentage)}
    \label{fig:q-results}
\end{figure*} 

\label{sec:strats}
The o-agent in the environment, will act using an optimal strategy until the x-agent is observed within the observation radius. Then,  adaptive learning strategies, RL and game-theoretic, will be used to make decisions and avoid collisions. 
\subsubsection*{Reinforcement Learning Algorithm}
The RL algorithm makes use of a neural network (NN) to find the optimal decision to take by treating the environment as a Markov decision process (MDP). The o-agent needs to find the optimal action $a_t$ to take while in state $s_t$ given its observation $o_t$ of its surroundings, which is done through Q-learning. This action is typically based on a policy $\pi$, which can be deterministic or stochastic. The goal of the RL algorithm is to use prior data to learn the optimal policy $\pi^*$ that maximizes the expected cumulative discounted reward. This uses a Q-function, represented as $Q^{\pi}(s_t,a_t)$, to determine how preferable it is to take action $a_t$ in state $s_t$. The Q-function is defined by the equation
\vspace{-0.1 in}
\begin{equation}
Q^{\pi}(s_t,a_t) = \mathbb{E}_{\pi}[\sum_{t=0}^{\infty}(\gamma)^tR_{t+1}|S_0 = s_t, A_0 = a_t]
\end{equation}

where $\gamma \in [0,1)$ represents the discount factor and $R_{t+1}$ represents the reward at time $t+1$. Using this formulation, the optimal policy $\pi^*$ is defined as 
\vspace{-0.1 in}
\begin{equation}
    \pi^*(s_t) = \arg \max_a (\max_\pi Q^{\pi}(s_t,a_t)).
\end{equation}

\subsubsection*{Game Theoretic Algorithm} 
The game theoretic algorithm uses the idea of a 2-player game between the o-agent and the x-agent within the observation radius. Each of the agents is competing with one another as they build a path to their destination. In this game, the o-agent is selecting an action that will maximize their own payoff. 

Let the action spaces for the x-agent and o-agent be $A^x$ and $A^o$ respectively. The payoff for agent $i$ is $u_i(a_i, a_{-i})$, where $a_{-i}$ represents the action of the other agent. Since the x-agent makes a decision following their own strategy, only the o-agent can be influenced. Therefore, the o-agent will find the Pareto optimal strategy, which is defined as taking the action $a^*$ for which there is no other $a$ such that:
\begin{equation}
    u_o(a) > u_o(a^*)
\end{equation}
Using the estimated trajectory outlined in \ref{subsec:estimation}, the o-agent can predict the most likely action for the x-agent and act accordingly to maximize the utility function. The utility function for the game theoretic approach can be described using
\begin{equation}
    u_o(a_o, a_i) = -\Delta d - \chi \phi_{col}
\end{equation}
where $\Delta d$ represents the change in distance to the destination after taking the action $a_o$, $\phi_{col}$ represents the potential for risk when taking the same action, and $\chi$ represents a weighting factor to determine how much to penalize a collision, which is chosed to be 10 in this work.



\subsection{Estimation}
\label{subsec:estimation}

To determine future actions to be taken by x-agents, an estimation algorithm is used by the o-agent. In this, historical observation data is used as an input to achieve a prediction for the next action to be taken by the x-agent. An approach to achieve this is using a neural network that can be trained with trajectories of agents moving through the environment. Equation \ref{eqn:input} shows how the input vector $s_{t,x_i}^{hist}$ is formed using $k$ previous observations of agent $x_i$ at time $t$ to predict its next action.
\begin{equation}
    s_{t,x_i}^{hist} = [(s_t,x_i), (s_{t-1,x_i}), ..., (s_{t-(k-1)},x_i)]
    \label{eqn:input}
\end{equation}

Using a NN also allows for a physics informed loss function, which allows the training to be constrained to the dynamics of the environment. The loss function would then be composed of two parts: mean squared error (MSE) loss and physics informed (PI) loss. An example loss function is shown in Equation \ref{eqn:loss}, where the PI and MSE loss components are shown in Equations \ref{eqn:pi-loss} and \ref{eqn:mse-loss} respectively. The variable $\hat{a}_{t+1,x_i}$ is the prediction coming out of the NN before adjusting the weights and $S_j$ represents each state in the set of possible states. In the basic-dynamics model utilized in this study, $S_j$ will represent the potential next state values for the agent. The lack of physical dynamics such as velocity and acceleration in this setup will render a PI loss term as not useful, causing the value of $\zeta$ to be 0.  However, more realistic simulation environment can can potentially benefit from it.
\begin{equation}
    \text{Loss Function} = \zeta f_{PI}(s_{t,x_i}^{hist}) + \delta f_{MSE}(s_{t,x_i}^{hist})
    \label{eqn:loss}
\end{equation}
\begin{equation}
    f_{PI} = \text{average}(\min_j||\hat{a}_{t+1,x_i} - S_j||)
    \label{eqn:pi-loss}
\end{equation}
\begin{equation}
    f_{MSE} = \sqrt{\text{average}(a_{t+1,x_i} - \hat{a}_{t+1,x_i})^2}
    \label{eqn:mse-loss}
\end{equation}

\subsection{Risk}
In a MAS, risk refers to the uncertainty in outcomes resulting from the interactions between autonomous agents in dynamic and often unpredictable environments. Each agent may have incomplete information, and their actions can affect both their own objectives and those of other agents. Accordingly, risk is defined as: 
\begin{equation}
    R = \sum_{i} P(E_i) \cdot S(E_i)
\end{equation}
where $P(E_i)$ and $S(E_i)$ represent the probability and severity of event $E_i$ respectively. A risk threshold can be calculated as shown in \cite{bhadriraju2021risk}, which indicates the maximum allowable risk.
\begin{figure}
    \centering
    \includegraphics[width=0.8\linewidth]{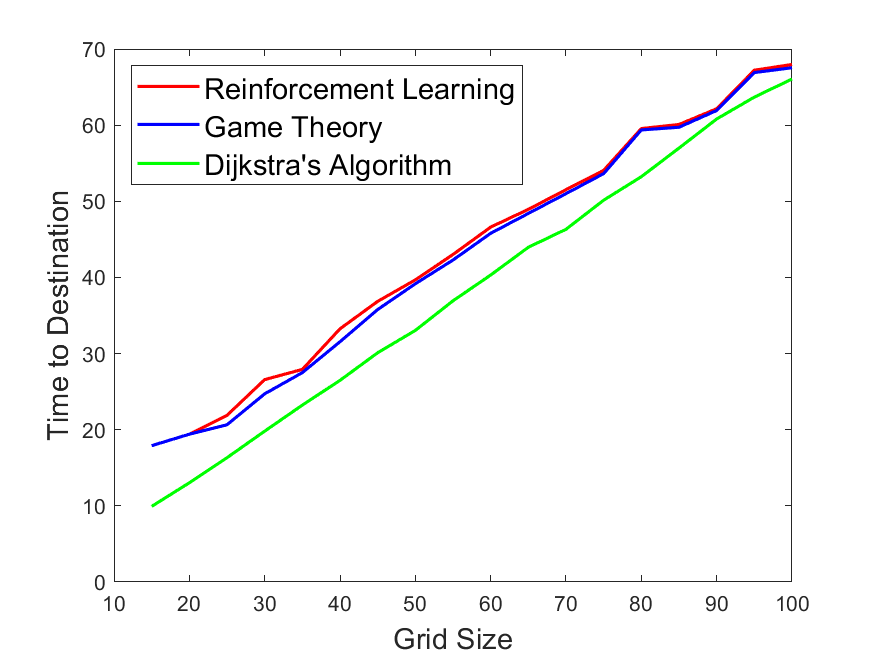}
    \caption{Comparison of learning algorithms and optimal strategy, with uses Dijkstra's algorithm, over varying grid sizes.}
    \label{fig:dijk-comp}
\end{figure} 

\section{Simulation and Results}
    \subsection{Environment Specifications}

 \begin{figure*}
     \centering
     \begin{subfigure}{0.40\textwidth}
        \centering
        \includegraphics[width=\linewidth]{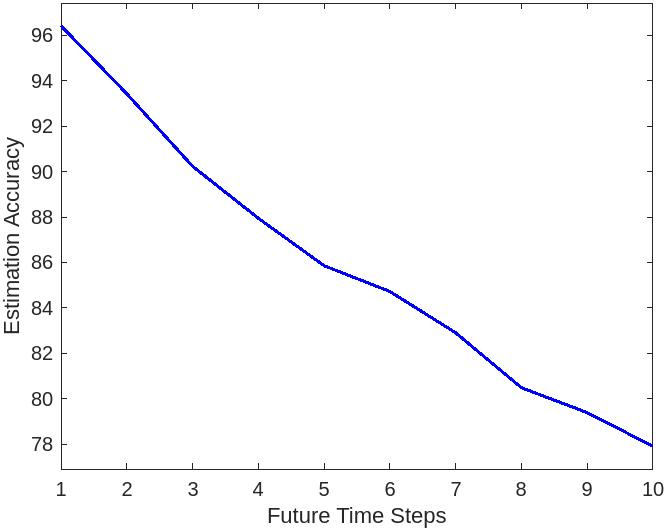}
        \caption{}
        \label{fig:acc-time}
    \end{subfigure}
    \hfill
    \begin{subfigure}{0.40\textwidth}
        \centering
        \includegraphics[width=\linewidth]{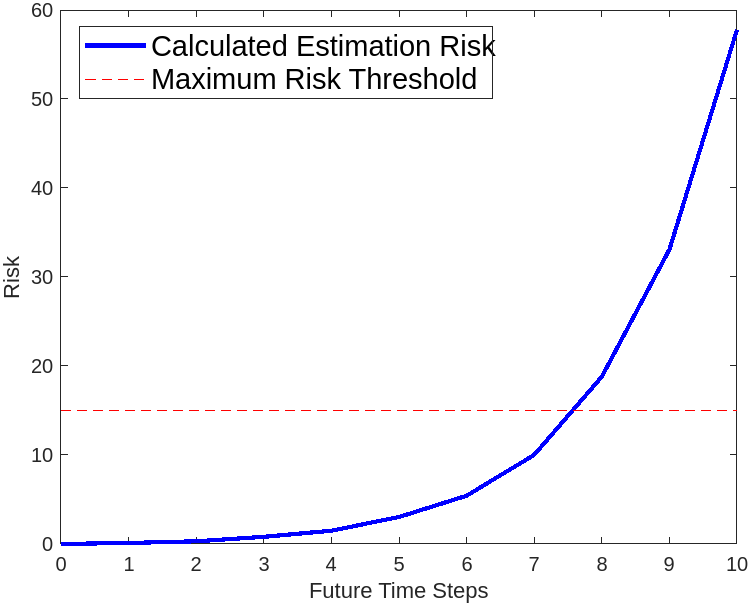}
        \caption{}
        \label{fig:risk-time}
    \end{subfigure}
     \caption{Results of the trained estimation algorithm: (a) prediction accuracy as a function of future time steps, (b) risk over future time steps}
     \label{fig:placeholder}
 \end{figure*}
The simulation was run on an AMD Ryzen 7 5700G, which is an x64-based processor with 16 GB of RAM and a 3.8 GHz clock speed. For software, the code was written in Python version 3.7 with access to the following libraries: numpy 1.26.4, scipy 1.13.1, tensorflow 2.18.0, pytorch 2.5.0, matplotlib 3.9.0, psutil 5.9.8. 

For the RL algorithm, the agent actions are decided through a neural network (NN) with input dimension matching the size of $s_{x_i}$ and output dimension matching the size of $a_{x_i}$. The NN contains two hidden layers, each with 128 nodes and was trained using Python's Tensorflow library. 

\subsection{Comparison of the Proposed Situation Awareness Framework using Two Learning Algorithms}
The study analyzed two learning algorithms by monitoring resource usage and performance metrics such as time to destination and number of collisions. Identical random environments were navigated using RL and game theory. Metrics were measured across grid sizes from 15 to 100 and observation radii from 0 to 7. Figure \ref{fig:q-results} depicts the average results of time to destination, number of collisions, memory, and CPU usage for both approaches.

The study highlights agent performance through time to destination and total collisions, while memory and CPU usage indicate resource consumption. Time to destination and total collisions remain relatively constant across different observation radius values, allowing agents to observe different sized areas and maintain similar performance. Comparing the resource consumption for differing values of observation radius is more difficult as the data is more inconsistent. However, there is a general trend showing higher memory and CPU usage for larger values of observation radius which is expected. Grid size significantly influences both performance and resource usage; time to destination appears to increase linearly with grid size, whereas total collisions, memory and CPU usage vary more. Smaller grids have high collision rates, whereas larger grids have fewer collisions as agents can space out. Larger grids use more memory but roughly equal CPU to smaller grid sizes. If avoiding collisions is crucial, a larger grid is better. In less collision-sensitive environments, smaller grids offer lower memory usage with similar CPU demand.

To add to the analysis, a study was done to compare how the time to destination for each learning algorithm compared to a strategy that took the shortest path. This is important to show how much the agent is willing to diverge from their course to avoid a potential collision. Since Figure \ref{fig:q-results} showed the observation radius having almost no dependence on the time to destination, this comparison only took place over a range of grid sizes from 15 to 100 in increments of 5. 

Data from identical trajectories using RL, game theory, and the optimal strategy (Dijkstra's algorithm) were analyzed. Over 100 iterations per grid size, using the same values as previously, the average destination time was calculated. Figure \ref{fig:dijk-comp} shows the optimal strategy grows linearly with grid size. Learning algorithms perform similarly across sizes and converge to the optimal strategy as grid size grows, indicating larger grids lead to solutions closer to optimal.



\subsection{Estimation}
\label{estimation}



An additional study was performed, where data was collected on the accuracy of estimation over multiple future time steps. This is achieved by using the NN iteratively where the output after the first time step becomes included in the input when calculating for the second time step. This procedure is repeated for up to ten time steps to understand how well the algorithm can maintain accuracy over time. The results are shown in Figure \ref{fig:acc-time}. As stated in Section \ref{subsec:estimation}, the loss function for the estimation NN is purely from MSE due to the basic dynamics of the environment.


Using this data, an additional study of risk over time steps was performed to observe how far into the future the algorithm can be relied upon. These results are shown in Figure \ref{fig:risk-time} and show that the estimation algorithm falls within the maximum risk threshold for seven time steps. 



\section{Conclusion}
    This study examined approaches for managing uncoordinated autonomous agents in multi-agent systems, comparing two learning algorithms for decision-making and varying observation radii for agent awareness. The algorithms were evaluated for performance and resource use, highlighting the trade-off between optimal strategy and computational complexity. An estimation algorithm was also developed to assess accuracy and risk over future time steps. Future work will include simulations in more realistic environments (e.g., 3D space), additional resource-constraints and adaptive strategies to balance effective performance with limited resources, and coordinated perception using multi-agent interactions and human-in-the-loop strategies.

\bibliographystyle{IEEEtran}
\bibliography{IEEEabrv,main}

\begin{thebibliography}{10}
\providecommand{\url}[1]{#1}
\csname url@samestyle\endcsname
\providecommand{\newblock}{\relax}
\providecommand{\bibinfo}[2]{#2}
\providecommand{\BIBentrySTDinterwordspacing}{\spaceskip=0pt\relax}
\providecommand{\BIBentryALTinterwordstretchfactor}{4}
\providecommand{\BIBentryALTinterwordspacing}{\spaceskip=\fontdimen2\font plus
\BIBentryALTinterwordstretchfactor\fontdimen3\font minus \fontdimen4\font\relax}
\providecommand{\BIBforeignlanguage}[2]{{%
\expandafter\ifx\csname l@#1\endcsname\relax
\typeout{** WARNING: IEEEtran.bst: No hyphenation pattern has been}%
\typeout{** loaded for the language `#1'. Using the pattern for}%
\typeout{** the default language instead.}%
\else
\language=\csname l@#1\endcsname
\fi
#2}}
\providecommand{\BIBdecl}{\relax}
\BIBdecl

\bibitem{zhang2023perception}
Y.~Zhang, A.~Carballo, H.~Yang, and K.~Takeda, ``Perception and sensing for autonomous vehicles under adverse weather conditions: A survey,'' \emph{ISPRS Journal of Photogrammetry and Remote Sensing}, vol. 196, pp. 146--177, 2023.

\bibitem{9827169}
W.~Jiang, X.~Xing, A.~Huang, and J.~Chen, ``Research on performance limitations of visual-based perception system for autonomous vehicle under severe weather conditions,'' in \emph{2022 IEEE Intelligent Vehicles Symposium (IV)}, 2022, pp. 586--593.

\bibitem{shimizu2021evaluation}
K.~Shimizu, D.~Suzuki, R.~Muramatsu, H.~Mori, T.~Nagatsuka, and T.~Matsumoto, ``Evaluation framework for performance limitation of autonomous systems under sensor attack,'' in \emph{Computer Safety, Reliability, and Security: 40th International Conference, SAFECOMP 2021, York, UK, September 8--10, 2021, Proceedings 40}.\hskip 1em plus 0.5em minus 0.4em\relax Springer, 2021, pp. 67--81.

\bibitem{9616449}
D.~Omeiza, H.~Webb, M.~Jirotka, and L.~Kunze, ``Explanations in autonomous driving: A survey,'' \emph{IEEE Transactions on Intelligent Transportation Systems}, vol.~23, no.~8, pp. 10\,142--10\,162, 2022.

\bibitem{BURTON2020103201}
\BIBentryALTinterwordspacing
S.~Burton, I.~Habli, T.~Lawton, J.~McDermid, P.~Morgan, and Z.~Porter, ``Mind the gaps: Assuring the safety of autonomous systems from an engineering, ethical, and legal perspective,'' \emph{Artificial Intelligence}, vol. 279, p. 103201, 2020. [Online]. Available: \url{https://www.sciencedirect.com/science/article/pii/S0004370219301109}
\BIBentrySTDinterwordspacing

\bibitem{de2020pysindy}
B.~M. de~Silva, K.~Champion, M.~Quade, J.-C. Loiseau, J.~N. Kutz, and S.~L. Brunton, ``Pysindy: a python package for the sparse identification of nonlinear dynamics from data,'' \emph{arXiv preprint arXiv:2004.08424}, 2020.

\bibitem{bhadriraju2021risk}
B.~Bhadriraju, J.~S.-I. Kwon, and F.~Khan, ``Risk-based fault prediction of chemical processes using operable adaptive sparse identification of systems (oasis),'' \emph{Computers \& Chemical Engineering}, vol. 152, p. 107378, 2021.

\bibitem{rosolia2018data}
U.~Rosolia, X.~Zhang, and F.~Borrelli, ``Data-driven predictive control for autonomous systems,'' \emph{Annual Review of Control, Robotics, and Autonomous Systems}, vol.~1, no.~1, pp. 259--286, 2018.

\bibitem{troullinos2021collaborative}
D.~Troullinos, G.~Chalkiadakis, I.~Papamichail, and M.~Papageorgiou, ``Collaborative multiagent decision making for lane-free autonomous driving,'' in \emph{Proceedings of the 20th International Conference on Autonomous Agents and MultiAgent Systems}, 2021, pp. 1335--1343.

\bibitem{chen2020deep}
C.~Chen, P.~Zhang, H.~Zhang, J.~Dai, Y.~Yi, H.~Zhang, and Y.~Zhang, ``Deep learning on computational-resource-limited platforms: A survey,'' \emph{Mobile Information Systems}, vol. 2020, no.~1, p. 8454327, 2020.

\bibitem{homberger2007multi}
J.~Homberger, ``A multi-agent system for the decentralized resource-constrained multi-project scheduling problem,'' \emph{International Transactions in Operational Research}, vol.~14, no.~6, pp. 565--589, 2007.

\bibitem{EATON1992705}
\BIBentryALTinterwordspacing
J.~W. Eaton and J.~B. Rawlings, ``Model-predictive control of chemical processes,'' \emph{Chemical Engineering Science}, vol.~47, no.~4, pp. 705--720, 1992. [Online]. Available: \url{https://www.sciencedirect.com/science/article/pii/000925099280263C}
\BIBentrySTDinterwordspacing

\bibitem{9180048}
P.~Karamanakos, E.~Liegmann, T.~Geyer, and R.~Kennel, ``Model predictive control of power electronic systems: Methods, results, and challenges,'' \emph{IEEE Open Journal of Industry Applications}, vol.~1, pp. 95--114, 2020.

\bibitem{holkar2010overview}
K.~Holkar and L.~M. Waghmare, ``An overview of model predictive control,'' \emph{International Journal of control and automation}, vol.~3, no.~4, pp. 47--63, 2010.

\bibitem{9361343}
G.~Torrente, E.~Kaufmann, P.~Föhn, and D.~Scaramuzza, ``Data-driven mpc for quadrotors,'' \emph{IEEE Robotics and Automation Letters}, vol.~6, no.~2, pp. 3769--3776, 2021.

\end{thebibliography}

\end{document}